\documentclass[preprint,journal]{vgtc}       

\onlineid{1797}

\vgtccategory{Research}

\vgtcpapertype{system}

\title{The Urban Toolkit:\\ A Grammar-based Framework for Urban Visual Analytics}

\author{%
Gustavo Moreira, Maryam Hosseini, Md Nafiul Alam Nipu,\\Marcos Lage, Nivan Ferreira, and Fabio Miranda
}

\authorfooter{
  \item
  	Gustavo Moreira, Md Nafiul Alam Nipu, and Fabio Miranda are with the University of Illinois Chicago.\\
  	E-mails: \{gmorei3,mnipu2,fabiom\}@uic.edu.
  \item
  	Maryam Hosseini is with the Massachusetts Institute of Technology.\\
  	E-mail: maryamh@mit.edu.
  \item Marcos Lage is with the Universidade Federal Fluminense.\\
  	E-mail: mlage@ic.uff.br.
   \item Nivan Ferreira is with the Universidade Federal de Pernambuco.\\
  	E-mail: nivan@cin.ufpe.br.
}

\abstract{
While cities around the world are looking for smart ways to use new advances in data collection, management, and analysis to address their problems, the complex nature of urban issues and the overwhelming amount of available data have posed significant challenges in translating these efforts into actionable insights. In the past few years, urban visual analytics tools have significantly helped tackle these challenges. 
When analyzing a feature of interest, an urban expert must transform, integrate, and visualize different thematic (e.g., sunlight access, demographic) and physical (e.g., buildings, street networks) data layers, oftentimes across multiple spatial and temporal scales.
However, integrating and analyzing these layers require expertise in different fields, increasing development time and effort.
This makes the entire visual data exploration and system implementation difficult for programmers and also sets a high entry barrier for urban experts outside of computer science.
With this in mind, in this paper, we present the Urban~Toolkit~(UTK), a flexible and extensible visualization framework that enables the easy authoring of web-based visualizations through a new high-level grammar specifically built with common urban use cases in mind.
In order to facilitate the integration and visualization of different urban data, we also propose the concept of \emph{knots} to merge thematic and physical urban layers.
We evaluate our approach through use cases and a series of interviews with experts and practitioners from different domains, including urban accessibility, urban planning, architecture, and climate science.
\highlight{UTK is available at \href{http://urbantk.org}{urbantk.org}.}
}


\teaser{
  \centering
  \includegraphics[width=1\linewidth]{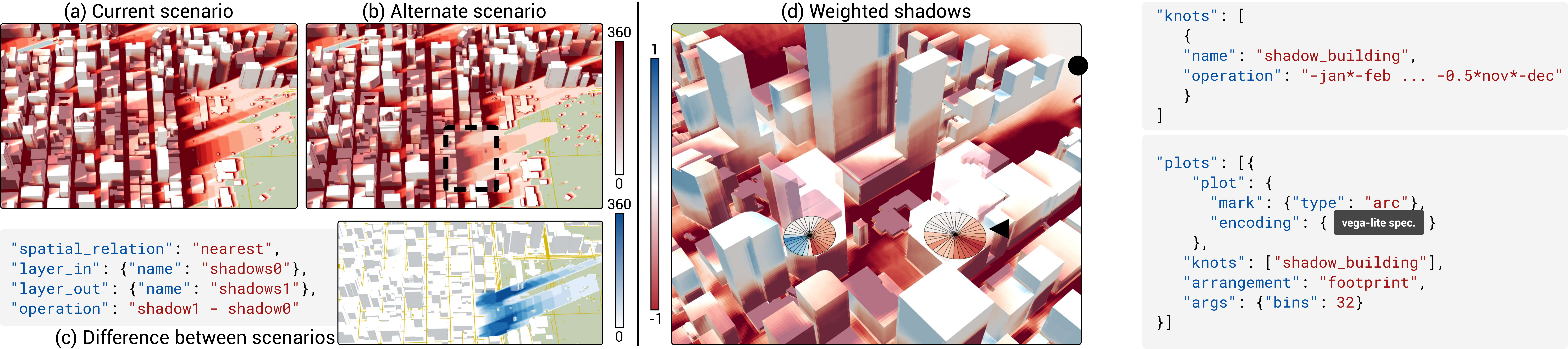}
  \caption{
  The Urban Toolkit enables the creation of urban visualizations through a new grammar that uses \emph{knots} to integrate thematic (e.g.,~sunlight access, demographic) and physical (e.g.,~buildings, street networks) data layers. Left: An example of \emph{what-if} scenario planning for urban shadow analysis. The difference between current (a) and proposed (b) urban development scenarios can be specified through algebraic operations using the grammar (c). Right: Sunlight access analysis at the building level. Chained operations are specified using the grammar, with plots embedded in the 3D environment (d).
  }
  \label{fig:teaser}
}

\graphicspath{{figs/}{figures/}{pictures/}{images/}{./}} 

\usepackage{mathptmx}                  
\usepackage{amsmath}
\usepackage{fixltx2e}

\usepackage{xspace}
\usepackage{amssymb}
\usepackage[normalem]{ulem}
\usepackage[varqu]{zi4}
\usepackage{helvet}

\begin{document}

\newcommand{\nivan}[1]{{\color{green} [Nivan: {#1}]}}
\newcommand{\fabio}[1]{{\color{red} [Fabio: {#1}]}}
\newcommand{\marcos}[1]{{\color{blue} [Marcos: {#1}]}}
\newcommand{\gustavo}[1]{{\color{orange} [Gustavo: {#1}]}}
\newcommand{\maryam}[1]{{\color{purple} [Maryam: {#1}]}}
\newcommand{\oursystem}[1]{UrbanTK\xspace}
\newcommand{\highlight}[1]{{#1}}

\setlength{\abovedisplayskip}{4pt}
\setlength{\belowdisplayskip}{4pt}

\definecolor{myblue}{HTML}{085579}
\definecolor{mypurple}{HTML}{795579}

\newcommand{\grammar}[1]{{\color{myblue}\fontfamily{zi4}\selectfont{#1}}}
\newenvironment{grammarenv}{\color{myblue}\fontfamily{zi4}\selectfont}{\par}
\newenvironment{apienv}{\color{mypurple}\fontfamily{zi4}\selectfont}{\par}



\firstsection{Introduction}
\maketitle

Cities around the world are looking for smart ways to make use of data to address their problems.
%
%
%
By analyzing data from cities, urban scientists, practitioners, policymakers, and communities can gain a deeper understanding of complex urban problems, enabling better planning, policies, and urban resilience, making government more efficient and, ultimately, improving the lives of residents.
But the complex nature of urban issues and the overwhelming amount of data have posed significant challenges in translating these efforts into actionable insights.
Drawing a comprehensive picture often involves collaboration across multiple disciplines, including computer science, engineering, climate sciences, architecture, urban planning, and public health, leading to a variety of analytical workflows and tasks to transform, integrate, and analyze different urban data layers, often across multiple spatial and temporal scales~\cite{doraiswamy_spatio-temporal_2018}.
%
%
%
Urban visual analytics is considered a key component in enabling such analyses. It has been used to augment data mining solutions and leverage domain knowledge~\cite{andrienko_interactive_2009, ferreira_visual_2013, ferreira_urbane_2015, chen_interactive_2016}, as well as public dissemination of findings and policies~\cite{malik_proactive_2014, hemmersam_exploring_2015, miranda_shadow_2019}.
While there have been many successful urban visual analytics tools, tackling a variety of urban problems~\cite{zheng_visual_2016,deng_survey_2023}, the area still suffers from a fragmented community and disconnected efforts \highlight{\cite{yap_free_2022}}.
Projects are often siloed and developed based on one-off collaborations between computer scientists and urban experts to tackle specific problems. 
In the end, new solutions are often built from scratch, leading to wasted efforts and results that are difficult to reuse and extend to different scenarios and geographical locations.
Furthermore, creating visual analytics prototypes and tools is still challenging and time-consuming, requiring expertise in visualization, computer graphics, and data management. 
This is especially true in the increasingly important case of 3D urban analytics~\cite{mota_comparison_2022}, in which the design and implementation of visualization systems are substantially more complex.
This results in a considerable effort even for experienced visualization developers, and also in high entry barriers for urban experts outside of computer science and communities lacking the necessary resources.

The visualization community has made significant strides in facilitating the authoring and prototyping of interactive visualization interfaces through toolkits and high-level grammars. While these contributions have significantly lowered the entry barrier for information~\cite{satyanarayan_vega-lite_2017, satyanarayan_reactive_2016, bostock_d_2011, zong_animated_2023, bostock_protovis_2009}, scientific~\cite{lyi_gosling_2022, rautek_vislang_2014, kindlmann_diderot_2016}, and immersive visualization~\cite{sicat_dxr_2019, butcher_vria_2021}, the same cannot be said about urban visual analytics. 
Urban data and urban analytical tasks impose specific requirements that must be met to drive real-world applications.
To democratize urban visual analytics, we need new capabilities that empower a broad range of stakeholders to analyze urban data and foster new reproducible, flexible, and extensible tools.
%
%

As a step towards this direction, in this paper, we propose the Urban Toolkit (UTK), an open-source visualization toolkit for the easy authoring and prototyping of urban visual analytics applications.
%
%
Considering such applications and tools, UTK proposes to enable the following: (1)~extensibility so that new functionalities by visualization researchers can be incorporated, (2) reproducibility, ease of use, and deployment by urban experts and communities, and (3) flexibility so that outcomes can be translated into different urban domains and geographical locations.
Our goal is to \highlight{minimize} the necessity of being familiar with low-level concepts (e.g., rendering) or coding (e.g., C++ or JavaScript) to enable prototyping, evaluation of visualization designs, and the creation of compelling and useful urban visual analytics tools, even for 3D analysis scenarios.
To achieve this, at the core of UTK is a high-level visualization grammar for creating and sharing map and plot-based interactive urban visualizations. 
The grammar uses \emph{knots} \highlight{(grammar-defined links, or ties, between spatial layers)} to easily integrate 2D and 3D data layers across multiple spatial resolutions.
UTK also offers a suite of functionalities to facilitate the ingestion and parsing of data from different sources, supporting a broad range of use cases.
%
%
A tight connection between the data handling and visualization capabilities enables the easy integration of UTK into existing urban workflows and the quick prototyping of visualization interfaces. 
%
\highlight{UTK is publicly available at \href{http://urbantk.org}{urbantk.org}.}

These functionalities can be seen from at least three different angles: first and foremost, as a unified framework aimed at urban scientists and experts that abstracts many low-level data management and visualization aspects involved in implementing urban visual analytics applications. Second, as a testbed for visualization researchers and practitioners interested in the implementation of new applications as well as the prototyping and user evaluation of design choices and techniques for urban visualization. And third, as an easy-to-use authoring tool for stakeholders and communities that enables the creation and sharing of interactive urban visualizations.
To highlight the usefulness of our proposal across diverse domains and workflows, we present a set of use cases that implement complex analysis scenarios using UTK. 
We also report on a series of interviews with urban experts from urban planning, architecture, and climate sciences.
The work reported in this paper can be summarized as follows:
\begin{itemize}[noitemsep,topsep=0pt,leftmargin=*]
    \item We introduce a high-level visualization grammar to facilitate the development and sharing of urban visual analytics applications.
    \item We conceptualize a flexible way to specify data integration across multiple spatial resolutions.
    \item We present the Urban Toolkit, a toolkit that facilitates the design, authoring, and deployment of urban visual analytics applications.
    \item We report on a set of use cases inspired by previous works demonstrating UTK's ease of use, flexibility, and reproducibility.
    \item We present the feedback obtained from domain experts, reporting their perspectives on UTK.
\end{itemize}

\section{Background}
\label{sec:background}




Urban environments are complex ecosystems where various agents, events, and infrastructures dynamically and continuously interact. That, coupled with the fast-paced rhythm of life, creates an intricate net of relationships, correlations, events, and phenomena that can be sensed, simulated, or modeled, generating a large-scale 
stream of data.
%
%
\highlight{Following previous works~\cite{chen_exploring_2017,mota_comparison_2022},} in this paper, we categorize the urban data analysis targets into \emph{physical} and \emph{thematic layers}. 
\highlight{Physical layers represent the city's physical aspects (e.g., buildings, road networks, parks, water bodies) as geometries.}
Thematic layers store the urban data from simulations, machine learning models, sensing initiatives, or surveys.
\highlight{Figure~\ref{fig:scales} shows a thematic layer with sunlight access data integrated with four different types of physical layers. Multivariate datasets might require the use of abstract visualizations~\cite{kraus_immersive_2022}.}
%

Traditionally, the decision-making process has relied on tools that mainly utilize certain urban datasets defined over 2D maps, using a flat city metaphor to represent the city environment~\cite{zheng_visual_2016}.
More recently, however, several works have explored alternatives that integrate the built and the natural environment into the analysis process~\cite{mota_comparison_2022}. 
The reason is that a large portion of urban data either relates to natural phenomena (e.g., wind~\cite{avini_wind_2019}, noise~\cite{tang_dynamic_2022}) or must be analyzed concurrently with physical layers (e.g., road topology~\cite{jiang_digital_2022}, buildings~\cite{redweik_3d_2017}, sidewalks~\cite{hosseini_citysurfaces_2022}). If this physical context is not considered, relevant insights can be lost~\cite{batty_visual_2014, doraiswamy_spatio-temporal_2018}. 
For example, civil engineers analyzing landslides or floods must integrate data from simulations and data representing buildings, mountains, and street networks to assess the impact of new disasters~\cite{cornel_interactive_2019}.
Urban planners, architects and community boards analyzing the impact of new buildings on surrounding ones must integrate view, sky exposure, and building data to inform design and neighborhood character~\cite{ferreira_urbane_2015}.
%
%
These analyses are often performed by teams with complementary objectives.
In summary, at the core of a growing number of urban analysis workflows is the fundamental integration between physical (including 3D) and thematic data layers over multiple spatial scales.

Designing and implementing interactive visualization systems for exploring multiple layers of data poses several challenges.
First, the design process of an urban visual analytics application is naturally an iterative one~\cite{sedlmair_design_2012}, but there have been few studies that guide the best practices in the field~\cite{neuville_3d_2019, mota_comparison_2022}.
As a consequence, multiple alternative designs have to be tried out before the final one is chosen.
Second, implementing visualizations and interactions, especially 3D ones, is not trivial. 
In fact, even for developers with experience in computer graphics technologies, a substantial effort is often necessary to iterate over different design choices.
Finally, while general GIS systems provide analytical capabilities alongside the possibility of defining dashboards, they are not extensible and have a long learning curve.
%
%
Therefore, a visual analytics framework that can support the exploration of urban data by different stakeholders can greatly benefit from a highly flexible approach.
Such a framework can allow the stakeholders to guide the analysis in any direction required without being constrained by the limitations of the system. That is especially relevant with urban data due to its intricacies and complex relationships. 
Moreover, it can serve as a test bed for evaluation studies, facilitating comparison among different visualization designs, integration schemes, and interactions.
Our overall goal in the present paper is to propose such a framework.

\section{Related work}
\label{sec:relatedWork}

In this section, we review prior work in two areas. First, we review prior work on urban visual analytics tools.
Then, we review prior work on grammars, toolkits, and authoring tools to facilitate the creation of new visualizations.

\subsection{Urban visual analytics}

Visual analytics systems are important tools for urban data analysis~~\cite{deng_survey_2023}.
However, these tools are often tailored to particular domains and designed to analyze specific data layers independently. 
%
For example, previous works focused on transportation and mobility~\cite{ferreira_visual_2013}, planning~\cite{miranda_urban_2017}, environmental pollution~\cite{deng_airvis_2019}, and public safety~\cite{garcia-zanabria_cripav_2022}.
More closely related to our work are tools that enable the integration and exploration of multiple urban data layers.
Chang et al.~\cite{remco_chang_legible_2007} proposed a system that juxtaposes a 3D view of the city with an information view that displays multidimensional urban data.
Ferreira et al.~\cite{ferreira_urbane_2015} proposed a visual analytics tool that integrates physical and thematic data layers for urban development analysis.
More recently, Ortner et al.~\cite{ortner_vis--ware_2017} and Miranda et al.~\cite{miranda_shadow_2019} proposed systems that integrate 3D spatial and non-spatial views for view impact and sunlight access analysis, respectively.
Zeng and Ye~\cite{zeng_vitalvizor_2018} proposed a system that integrates physical entities and design metrics to study urban vitality.
\highlight{Sun et al.~\cite{sun_embedding_2017} proposed to embed visualizations into the physical layer, while Speckmann and Verbeek~\cite{speckmann_necklace_2010} and Angelini et al.~\cite{angelini_surgerycuts_2019} proposed map distortion techniques, with the latter inspired by manipulations of volumetric data~\cite{correa_feature_2006}.}
The integration of layers is also explored in tools for disaster management~\cite{cornel_visualization_2015,vuckovic_combining_2021}.


While these tools have shown valuable contributions in tackling various urban problems, the vast majority is not publicly available. This is a serious limiting factor for urban experts interested in using these tools across different domains or regions. Moreover, even if they were available, many rely on complex datasets that require laborious and time-consuming steps to parse and clean. 
%
Popular commercial and open-source GIS tools, while providing spatial analysis capabilities and abstracting many of these challenges, do so to the detriment of flexibility. They also offer limited support for the integration of spatial and abstract visualization components, as well as the integration of multiple physical and thematic layers.

UTK shares many similarities with previous efforts in urban visual analytics. It offers a unified method to integrate diverse data layers and supports multiscale analysis, as well as spatial (including 3D) and abstract visualization components. 
By adopting a declarative visualization grammar and streamlining data integration computations, we hope to facilitate the use and deployment of applications by urban stakeholders. We also hope to offer flexibility to visualization researchers so that they can extend and adapt the system as needed while ensuring reproducibility of outcomes to enable sharing of results across different domains and regions.



\subsection{Grammars and authoring tools for visualization}

%
%
Urban visual analytics tools are systems specifically built to gain insight into datasets and facilitate domain use cases.
Because of their complexity and design process, these tools are usually difficult to adapt to other scenarios, datasets, regions, or scales.
To facilitate deployment, flexibility, and reproducibility, general authoring tools have also been adopted by urban experts, enabling the creation of visualization interfaces, sometimes even removing the requirement to know how to program.
These works can be divided into three categories: template-based tools that expose their functionalities through GUIs, low-level libraries requiring programming knowledge, and high-level visualization grammars.
Given the ubiquity of spatial data, many of these tools support the creation of maps-based visualizations. 

In the first category, we have general template-based tools that follow a drag-and-drop metaphor for authoring visualization interfaces~\cite{noauthor_tableau_nodate,wongsuphasawat_voyager_2017,noauthor_trifacta_nodate}. 
While enabling the easy creation of visualization interfaces, they are too general to support the complex data and analytical tasks of real-world urban applications~\cite{zheng_visual_2016}. Their mapping capabilities are also restricted to 2D spatial regions.
%
%
In the second category, we have low-level visualization libraries~\cite{bostock_d_2011,noauthor_bokeh_nodate,noauthor_matplotlib_nodate}. These libraries offer some facilities for the creation of well-known charts, as well as the freedom to create new visualization designs. deck.gl~\cite{noauthor_deckgl_nodate}, for example, offers functionalities to create layer-based spatial visualizations. However, it lacks support to integrate, interact and link multiple layers and visualizations.

In the third category, we have high-level visualization grammars~\cite{satyanarayan_vega-lite_2017, park_atom_2018, sicat_dxr_2019,  shih_declarative_2019, liu_atlas_2021, kim_cicero_2022, zong_animated_2023}, a compromise between the ease of use of template-based tools and the flexibility of visualization libraries. Rather than be constrained by templates or requiring the programming of individual visualization components, visualization grammars empower users to specify their visualizations through high-level abstractions. In doing so, visualization specifications and system components are clearly separated.
Given their ease of use and flexibility, these grammars allow for quickly iterating over different visualization designs.
For example, with Vega~\cite{satyanarayan_reactive_2016}, Vega-Lite~\cite{satyanarayan_vega-lite_2017}, and Animated Vega-Lite~\cite{zong_animated_2023}, users can author their own visualizations through JSON files following rules that specify marks, encodings, and interactions of the plots.
DXR~\cite{sicat_dxr_2019}, VRIA~\cite{butcher_vria_2021}, and Deimos~\cite{lee_deimos_2023} extend Vega-Lite's grammar to virtual and augmented reality, offering the ability to create immersive visualizations within arbitrary (including urban) environments, but without offering complex integration and interaction between different data layers.

Our work is inspired by and extends the aforementioned high-level visualization grammars. 
UTK, beyond offering a unified and streamlined framework to load and parse urban layers, also provides a conceptual model to integrate physical and thematic layers across multiple resolutions. This model is made available to the user through a declarative grammar, enabling data transformation while abstracting low-level implementation details.

\begin{figure*}[h!]
    \centering
    \includegraphics[width=0.95\linewidth]{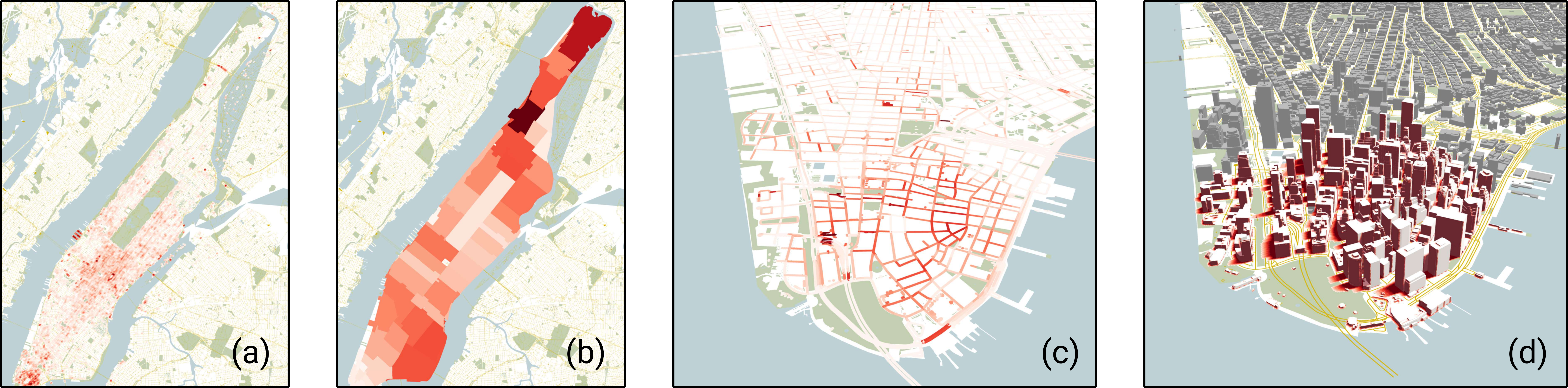}
    \caption{UTK supports the integration of thematic and physical layers at multiple scales through grammar-defined knots. For example, sunlight access aggregated at a grid (a), ZIP code (b), and street level (c). It also supports thematic data over building surfaces (d).}
    \vspace{-0.5cm}
    \label{fig:scales}
\end{figure*}

\section{UTK design goals}
\label{sec:design}

In this section, we synthesize the design goals that guided the development of UTK.
These principles were motivated by our previous contributions and collaborations with urban experts~\cite{ferreira_urbane_2015, doraiswamy_interactive_2018, miranda_shadow_2019, mota_comparison_2022, hosseini_citysurfaces_2022} and also by our goal of making the general tasks of designing, prototyping and sharing interactive urban visualizations easier.
Recurrent meetings with an urban expert (co-author of the paper) also influenced the design and implementation of the toolkit.
\highlight{
These meetings can be divided into three phases.
In the initial meetings, we focused on establishing a shared vocabulary and on critically reflecting on the limitations and tradeoffs of previous works.
Then, we reflected on our experience as visualization researchers, tool builders, and urban experts and how our practices could be improved with a unified framework. This process led to a set of design goals and capabilities for UTK.
In the final phase, as the framework took shape, we reflected on whether the design goals were being met.}
%
%
\highlight{Each broad design goal is divided into fine-grained goals.}
In the follow-up sections, we refer back to these goals when describing UTK's grammar (Section~\ref{sec:grammar}) and overall framework (Section~\ref{sec:utk}). Section~\ref{sec:reflections} reflects on the design goals.





\noindent \textbf{(D1) Extensibility.}
The extensibility to create new visualization designs and use these designs at different scales and aggregations is paramount to urban visual analytics tools.
Visual analysis of data happens through an iterative process where insights are roughly derived through free exploration and formulation of hypotheses.
This freedom is impacted by the system's range of features and execution possibilities and the amount of \emph{friction} it generates during its use.
The smaller the gap between what the user conceptualizes as the next exploratory step and the actualization of it, the smaller the friction~\cite{zhicheng_liu_mental_2010}.
Minimizing this gap is not trivial, especially considering complex data such as urban ones, and the relationship between multiple data layers.
With this in mind, it is important to offer the user the \textbf{ability to create custom visualizations} that match, to the best extent possible, their exploration model (\highlight{\textbf{D1.1}}).
These models might also \textbf{include complex operations, such as aggregations or comparisons} (e.g., computing the difference between two or more development scenarios~\cite{delaney_visualization_2000, chakraborty_scenario_2015}) (\highlight{\textbf{D1.2}}).
It is also important to \textbf{abstract low-level tasks} (loading data, rendering, interaction, navigation) so that the user can focus on the actual tasks that more directly impact the visual analytics process (\highlight{\textbf{D1.3}}).

\noindent \textbf{(D2) Reproducibility.}
Reproducing an urban visual analytics system is neither an easy nor sufficiently discussed task. Most of the previous works remain closed-source projects, to the detriment not only of the broader visualization community but also of urban experts.
These systems are often a complex amalgamation of features, techniques, models, and libraries. Therefore, even though the data is readily available in many cases, the development of the surrounding system is time-consuming and full of intricate implementation details.
Rather, we should strive to enable easier collaboration between users and increase the verifiability of future research, especially when considering that many of these tools aim to influence decision-making in real-world settings~\cite{shi_integrating_2021}.
We should increase reproducibility in two major ways: (1) \textbf{Standardizing formats and inputs} so that data can be easily packaged and shared (\highlight{\textbf{D2.1}}), and (2) \textbf{Treating the entire system as an artifact that is shaped around a high-level specification} defined by the user (\highlight{\textbf{D2.2}}).
Since the system is built from a specification, reproducing it should be as simple as sharing a JSON file with supplementary datasets.


\noindent \textbf{(D3) Flexibility.}
The principle of how flexible a tool is to be adapted to other regions, scales, and data is also rarely considered.
Most visual analytics contributions follow one-off collaborations with urban experts and result in highly specialized tools to a single scenario.
This flexibility is seldom listed as a requirement for these tools. 
From a visualization perspective, the effort needed to re-engineer a tool to work in another region or with other data might not be worthwhile if the proper incentives are not in place (e.g., collaborators to use the tool, perspectives towards a new publication).
From an urban domain perspective, extending tools often requires expertise in highly specialized areas of computer science beyond what can be expected from an urban expert.
The status quo is where urban visual analytics tools are rarely adopted by experts beyond the original collaborators that helped in the design process.
A well-designed framework must offer \textbf{support to changing region, scale, or data} (\highlight{\textbf{D3.1}}) and offer solutions to \textbf{minimize friction and retooling while avoiding exposing low-level functionalities to the user} (\highlight{\textbf{D3.2}}).

\begin{figure*}
    \centering
    \includegraphics[width=1\textwidth,keepaspectratio=true]{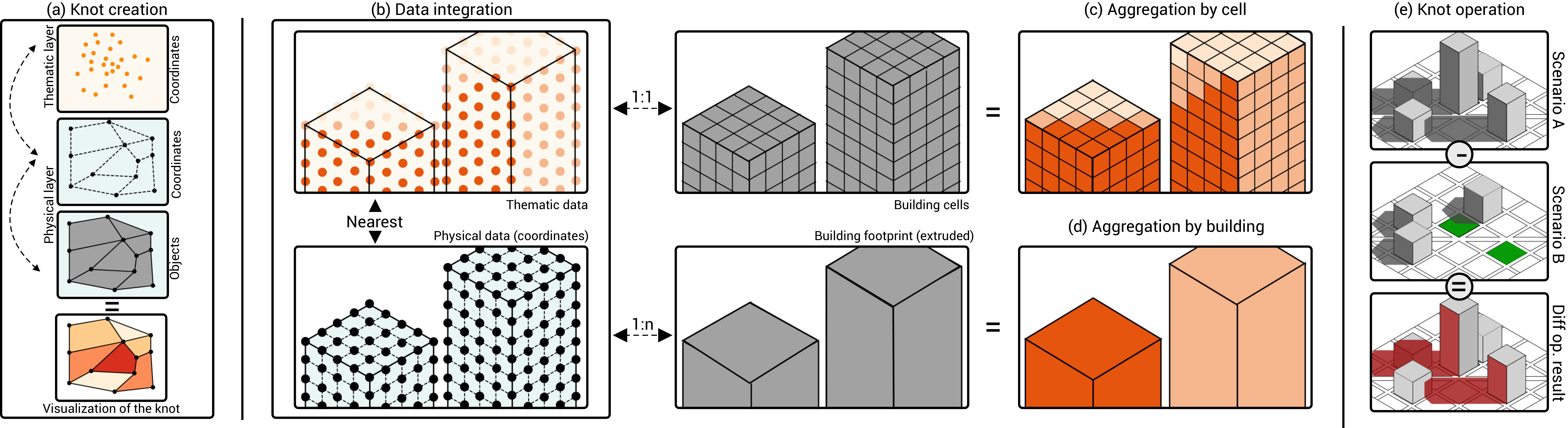}
    \caption{Integration between thematic and physical layers. (a) A knot is defined as the combination of a thematic and a physical layer through a spatial relation (e.g., nearest, contains) and an aggregation operator (e.g., sum, mean) if the relationship is $1:n$. Each thematic data point will be mapped to an individual physical layer coordinate (e.g., grid) and/or object. (b) For example, thematic data that is defined at a regular grid can be linked with a grid defined at the building level. Aggregations can then be performed at a building cell level (c) or building footprint level (d). (e) The concept of knots also allows for more complex data operations, such as the comparison between alternate planning scenarios.}
    \vspace{-0.5cm}
    \label{fig:integration}
\end{figure*}

\section{A visualization grammar for urban visual analytics}
\label{sec:grammar}

With the aforementioned design principles in mind, we now present the key concept behind UTK: a visualization grammar specifically designed for urban visual analytics.
The grammar allows for a detailed and precise specification of the entire visualization interface:
views, camera position and direction (for spatial views), data layers (including operations on top of them), and plots.
The former is the primary component of the grammar, and all other items are defined in the context of a view.
That being said, a view can be manipulated as any other item and can be used to compose multiple or embedded views.
Next, we detail our grammar's rules, grouping them into three categories: composing views, data layers (including thematic, physical, layer integration, and operation), and plotting data.
\highlight{For our abstract code description, we follow Ren et al.'s notation~\cite{ren_charticulator_2019}, with \grammar{:=} representing assignments, \grammar{|} representing ``or'', \grammar{?} representing optional elements, \grammar{+} representing one or more elements, and \grammar{*} zero or more.}

%


%

\subsection{Composing views}
\label{sec:composing_views}

At the highest level of our visualization grammar, we have the element \grammar{views} containing one or more \grammar{(view, camera)} elements of the visualization.
Each individual \grammar{view} element is defined by:
\highlight{a \grammar{map}; \grammar{knots}, defining links between data layers (Section~\ref{sec:data_layers}); and optional \grammar{plot}s, defining 2D plots (Section~\ref{sec:plotting_data}).
\highlight{Knots store the integrated data that will be visualized in both maps and plots.}
Considering Figure~\ref{fig:teaser}~(d), \grammar{map} will define the overall 3D map, \grammar{knots} will define the integration between physical (i.e., buildings) and thematic (i.e., shadow) layers, and \grammar{plot} will define the embedded radial plot.
}
%
%
\begin{grammarenv}
\begin{align*} 
\text{views} &:= (\text{view}^+, \text{camera}^+)\\
\text{view} &:= (\text{map}, \text{knots}, \text{plot}^*)\\
\text{map} &:= (\text{camera\_id}, (\text{knot\_id}, \text{interaction})^+)\\
\text{camera} &:= (\text{camera\_id}, \text{position}, \text{direction})\\
\text{interaction} &:= \text{brush}\ |\ \text{pick}
\end{align*}
\end{grammarenv}

A combination of multiple \grammar{view} elements should enable the creation of flexible juxtaposed visualizations~\cite{mota_comparison_2022}. If a \grammar{camera} is shared across \grammar{view}, then the same camera transformations are used in the views. Conversely, if different \grammar{camera} elements are used, we have a multi-view interface.
%
Finally, a \grammar{map} is defined as reference to a camera and a list of tuples with a reference to a \grammar{knot} to be rendered and an \grammar{interaction}.
The brushing interaction allows users to select a subset of a layer.
On the other hand, the picking interaction allows the user to select entire objects at once (considering the definition of an object at the geometry level).

\subsection{Specifying data layers}
\label{sec:data_layers}
The capacity to load, integrate, visualize, and perform operations on top of data layers sets UTK's grammar apart.
As previously mentioned, UTK organizes layers into two types: \emph{thematic} and \emph{physical} layers.
Thematic data layers correspond to discrete measurements over the 2D or 3D space. For example, noise complaints and crime occurrences over 2D regions or shadow values over 3D building surfaces.
Physical layers correspond to the built or natural environment, such as buildings, road networks, mountains, or regions of interest in a city, such as neighborhoods and parks.
The physical layer will be responsible for defining all the geometric information that will ultimately be rendered on the map, while the thematic layer carries the data attributes that will be linked to the physical layer. 
%
Thematic layers must be defined through a set of discrete points (coordinates). 
Physical layers, however, can only be defined through a set of objects and their coordinates:
\begin{grammarenv}
\begin{align*}
\text{layer}\textsubscript{thematic} &:= \text{(layer\_name,coordinates,color\_scale)}\\
\text{layer}\textsubscript{physical} &:= \text{(layer\_name,objects,coordinates)}
\end{align*}
\end{grammarenv}

%
A \grammar{color\_scale} will map the thematic data domain to a color range.
\highlight{Both thematic and physical layers are composed of points}.
For the physical layers, these coordinates can be grouped together to form a 2D or 3D polygon (object), such as ZIP area.
%
%
A key aspect in many urban analytical workflows is the necessity to navigate and visualize data at multiple scales (Figure~\ref{fig:scales}).
The separation of data through layers allows us to cover these scales by simply specifying aggregations of thematic layers over physical layers.
For example, a macro-scale analysis can be performed by joining a thematic layer with a physical layer describing neighborhoods.
Similarly, a meso-scale analysis can be performed with a join between the thematic layer and a physical layer describing lots and a micro-scale analysis with a physical layer describing buildings.

\subsubsection{Integrating layers}

UTK's grammar allows for spatial joins to be performed through \emph{knots} -- i.e., an explicit integration between thematic and physical layers.
Knots are high-level data integration operations that link thematic and physical layers, and each \grammar{view} is composed of a series of \grammar{knots}.
\begin{grammarenv}
\begin{align*}
\text{knots} &:= (\text{knot} \otimes \text{filter})^+\\
\text{knot} &:= (\text{knot\_name},(\text{integration\_scheme})^+)\\
\text{filter} &:= \text{bounding\_box} \text{ | } \text{address}
\end{align*}
\end{grammarenv}

Each \grammar{knot} is defined as a series of chained \grammar{integration\_scheme} elements.
An \grammar{integration\_scheme} is defined as a \grammar{spatial\_relation} (e.g., nearest, contains, within) between either two knots or two layers \highlight{(specified by their names)}:
\begin{grammarenv}
\begin{align*}
\text{integration\_scheme} &:= ((\text{layer}\textsubscript{in} \text{|} \text{knot}\textsubscript{in}),
(\text{layer}\textsubscript{out} \text{|} \text{knot}\textsubscript{out}),\\
&\hspace{15pt}\text{spatial\_relation?}, \text{operation?})\\
\text{spatial\_relation} &:= \text{nearest} \text{ | } \text{contains ...}
\end{align*}
%
\end{grammarenv}

\highlight{A filter \grammar{$\otimes$} operation performs a selection of items from knots taking into account a bounding box or specified address.}
%
%
If the spatial relation is $1:n$ (i.e., physical object contains $n$ thematic data points), we can specify an aggregation \grammar{operation} (e.g., mean), \highlight{or a user-defined JavaScript function (\grammar{custom\_function}) that aggregates the values}.
\begin{grammarenv}
\begin{align*}
\text{operation} &:= \text{aggregation | custom\_function}\\
\text{aggregation} &:= \text{ min | max | sum | mean ...}
\end{align*}
%
\end{grammarenv}

\begin{figure}[b!]
    \centering
    \includegraphics[width=1\linewidth]{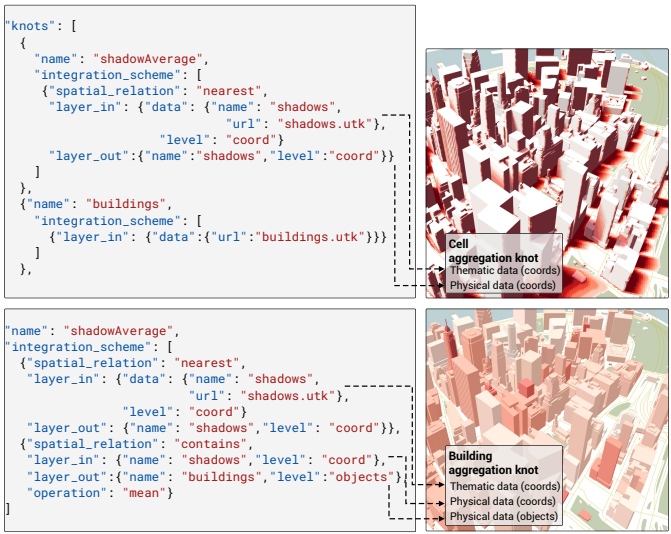}
    \caption{UTK abstracts spatial joins through the use of \emph{knots}. Top: Knot integrating thematic and physical layers. Bottom: Knot integrating a thematic layer with the average per building.}
    \label{fig:knots}
\end{figure}

Knots provide a flexible and intuitive way to chain integration schemes over multiple physical scales.
For example, thematic point data (e.g., taxi pickups) can be aggregated over physical object data (e.g., ZIP codes) by specifying a \grammar{knot} with \grammar{contains} (i.e., a link between point and object) followed by an \grammar{aggregation}.
This concept also allows for more complex integration, such as the ones considering sunlight access values defined at the surface of a building. For example, consider the case shown in Figure~\ref{fig:integration}~(b,~c).
\highlight{In (b,~top), an input thematic layer contains sunlight access values defined at a coordinate level. A nearest \grammar{spatial\_relation} creates a link between each thematic coordinate and the nearest coordinate in the output building physical layer (\highlight{b,~bottom}).
Aggregations can then be performed at building cell level (c) or footprint level (d).
Figure~\ref{fig:knots} shows the results of these aggregations.
}
%
%
%
\highlight{Through these operations, UTK supports multi-scale exploration (\textbf{D3.1})}.
%

\begin{figure*}[th!]
    \centering
    \includegraphics[width=1\linewidth]{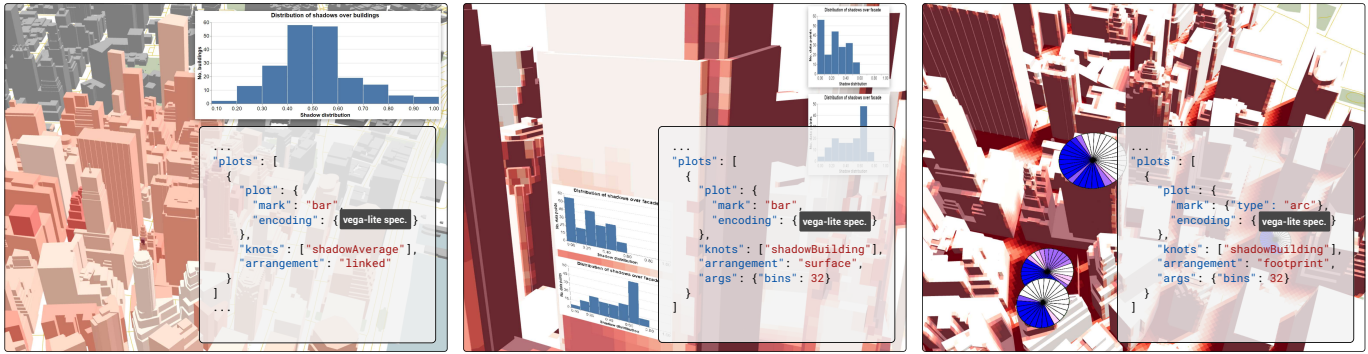}
    \caption{Different types of physical and thematic layers integration. Left: \grammar{linked} view showing the distribution of sunlight access. Middle: \grammar{embedded} \grammar{surface} plots showing the distribution of sunlight access over building surfaces. Right: \grammar{embedded} \grammar{footprint} plots showing horizontal cross-section distribution of sunlight access.}
    \vspace{-0.5cm}
    \label{fig:plots}
\end{figure*}

\subsubsection{Operations with knots}

Knots also facilitate the specification of layer operations, \highlight{supporting \textbf{D1.2}}. This is particularly important for \emph{what-if} analyses in which an urban expert must compare different built environment configurations and evaluate the one that best satisfies certain criteria (e.g., building designs that minimize the amount of shadow on neighboring parks).
Figure~\ref{fig:teaser}~(left) presents an example of the shadow in two scenarios: with and without two buildings near a park in Chicago. The difference (Figure~\ref{fig:teaser}~(c)) is shown in shades of blue.
UTK's grammar enables such operations to be specified through the \grammar{operation} element. For example, the knot \grammar{(difference, (nearest, shadow$_0$, shadow$_1$, shadow$_0$-shadow$_1$))} computes the difference between two knots (\grammar{shadow$_0$} and \grammar{shadow$_1$}). Note that since both knots are defined over the same physical layer (i.e., buildings), the \grammar{nearest} element is redundant.
\highlight{Operations can only be performed between knots with the same physical layer, but their thematic layers can have different resolutions. Since knots will map each thematic data point to an individual physical layer coordinate (through the \grammar{integration\_scheme}), this guarantees that there will not be any mismatch between knots.}
%

\subsection{Plotting data}
\label{sec:plotting_data}

UTK's grammar also facilitates the creation of 2D plots, both linked to the view or embedded in the 3D environment. Both types have been shown to be useful in our previous evaluations~\cite{mota_comparison_2022}, \highlight{and support \textbf{D1.1}}.
For the specification of these plots, we enable the insertion of Vega-Lite's specifications into one defined according to UTK's grammar. In doing so, we leverage a wealth of easy-to-use visualizations that have been proposed by the Vega community at large (\textbf{D1}).
A UTK plot is defined~as:
\begin{grammarenv}
\begin{align*}
\text{plot} &:= \text{(vega\_spec,(knot\_name,arrangement)$^+$,}\\
&\hspace{15pt}\text{interaction?,args?)}\\
\text{arrangement} &:= \text{linked | embedded}\\ 
\text{embedded} &:= \text{surface | footprint ...}
\end{align*}
%
\end{grammarenv}

\noindent where \highlight{\grammar{vega\_spec} is a Vega-Lite specification.}
Two types of arrangements are supported. In a \grammar{linked} arrangement, plots are displayed on top of the 3D map. In an \grammar{embedded} arrangement, plots are embedded directly into the 3D map. Two different embedding arrangements are supported: \grammar{surface}, where a 2D plot is displayed on the surface of a physical layer, and \grammar{footprint}, where a 2D plot is displayed on the horizontal cross-section of a physical layer.
Plot parameters (e.g., number of segments in a radial plot) are also defined in the grammar (\grammar{args}) and passed to both Vega-Lite and to the frontend (to compute the intersection between building and horizontal cross-section).
Figure~\ref{fig:plots} shows examples of these possibilities.

\section{UTK: A toolkit for urban analytics}
\label{sec:utk}

UTK is a general toolkit for urban analytics that uses the previously introduced visualization grammar to create new map-based visualizations.
However, given the complexity of urban data and workflow, UTK also offers data loading and parsing capabilities.
Our framework can then be divided into two modules, detailed next.

\subsection{Backend module}

UTK's backend is a collection of data management functionalities for \highlight{loading, saving, transforming, and aggregating the data into formats appropriate to the frontend.}
Its functionalities are exposed through a Python API that can be accessed in computational notebooks.
\highlight{In line with \textbf{D1.3}}, many of the time-consuming aspects of urban analytics are abstracted by this component in such a way that, in order to have the data for an initial urban interface, the user only needs to:

\begin{apienv}
\noindent import utk\\
uc = utk.OSM.load(`Chicago,USA',layers=[`buildings'])\\
uc.save(`chi')\\
uc.view()
\end{apienv}

This will download and parse OpenStreetMap's (OSM's) building data for the city of Chicago and open a web browser with a default grammar specification to visualize the data.
UTK supports a wide range of data that model physical and thematic aspects of a city.
To ensure consistency, the physical layer is immutable in the system, meaning that all operations done using knots only transform the thematic layer. The immutability of the physical layers adds stability to the system since the users do not need to be concerned about ensuring that the operations maintain the mapping of the physical/geometric representations.
The layers can be loaded from multiple data formats, including Pandas' DataFrames, GeoJSONs, shapefiles, and CSV files, supporting \textbf{D2.1}.

\noindent \textbf{Physical layers.}
We support four types of physical layers:

\noindent \emph{OpenStreetMap data.}
OSM provides spatial data for a large number of cities. 
The data, however, is specified as tags and elements that represent a number of urban features, such as parks, water bodies, and buildings. Such data must be parsed to be used by rendering and visualization pipelines.
With UTK, a user can specify a region of interest through a bounding box, polygon, or address (similar to the example above).
UTK will automatically download and parse the data into UTK-ready layers and formats, such as 3D triangle meshes for OSM buildings, 2D polygons for OSM parks, and network data for street networks.
The framework also supports the Protocolbuffer Binary Format -- a data format used to store OSM data locally.

\noindent \emph{Polygons and triangle meshes.}
UTK supports both 2D polygons as well as 3D triangle meshes representing physical features, such as neighborhoods, ZIP codes, and buildings.

\noindent \emph{Network data.}
Our framework also supports street and footpath sidewalk data that usually describes a graph where vertices correspond to street corners and edges to road or sidewalk segments.

\noindent \emph{Grid.}
UTK supports a fine-grained grid that covers the city and/or the surface of buildings. These grids can be used to aggregate and overlay thematic data over surfaces.



\noindent \textbf{Thematic layers.}
UTK can load data where each point is defined by a latitude, longitude, height, and value tuple.
In practice, any value attributed to a coordinate in the 3D space can be loaded.

As an example to showcase UTK's flexibility in using data from previous efforts, we have incorporated Miranda et al.~\cite{miranda_shadow_2019} shadow accumulation algorithm \highlight{and also made it available via our Python API.}
%
%
%
For each surface point in the urban environment, the computed dataset contains the accumulated shadow over a user-specified period of time -- i.e., how much of the view to the sun was occluded considering the accumulation period.
That being said, all data used in the grammar specification must follow a two-step loading process. The first step involves preprocessing the supported data so that they conform to a format accepted by UTK.
The second step establishes links between the different layers to create knots, as specified in the grammar. \highlight{The .utk files are the result of this two-step process.}

\subsection{Frontend module}

\begin{figure}[t!]
    \centering
    \includegraphics[width=0.95\linewidth]{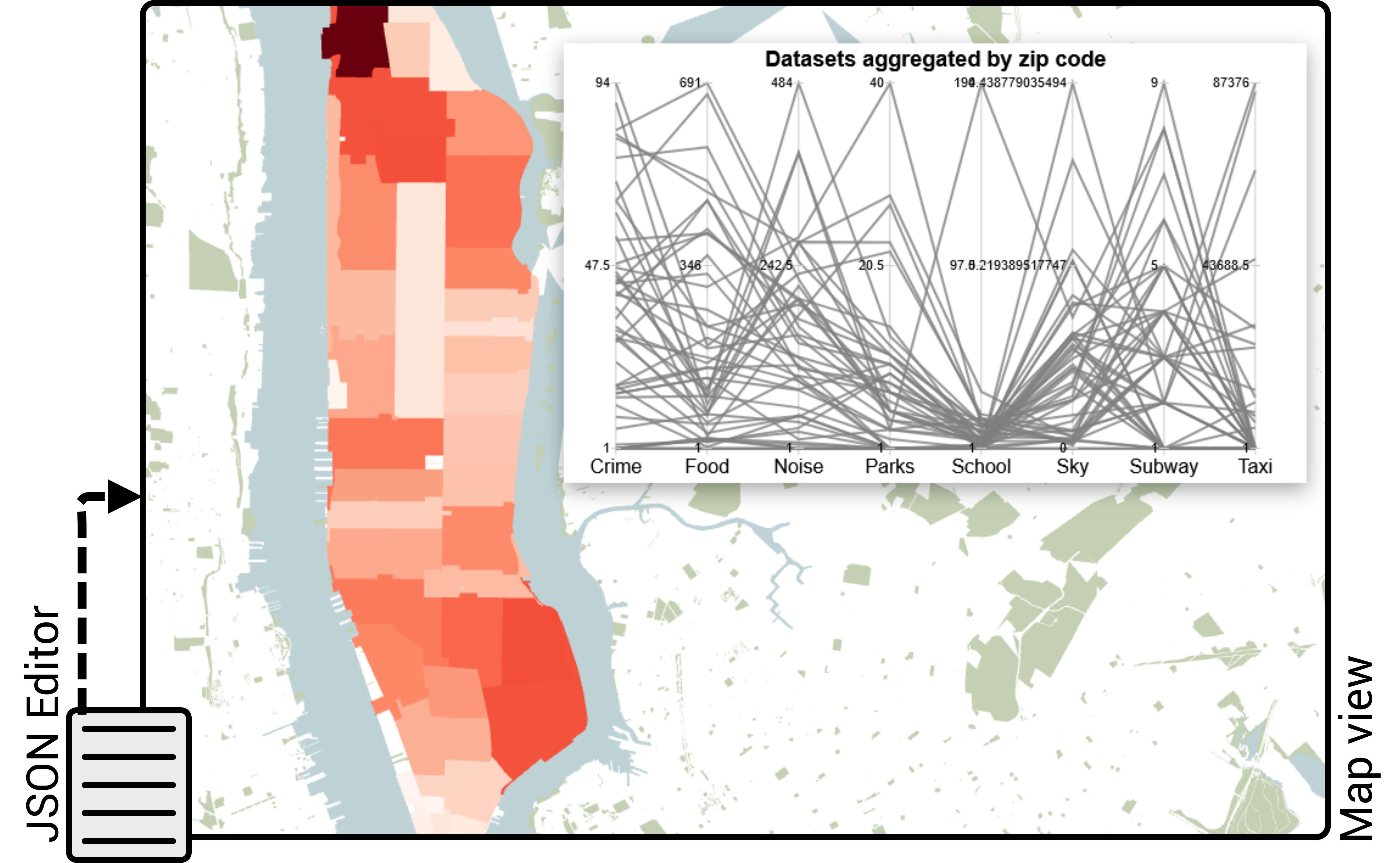}
    \caption{\highlight{UTK's frontend has a built-in JSON editor, and modifications to the JSON file are visualized in the map view.} Here, we visualize the spatial distribution of noise complaints in Manhattan, with a linked parallel coordinate plot showing other urban datasets.}
    \vspace{-0.5cm}
    \label{fig:interface}
\end{figure}

UTK's web interface is composed of two components: A JSON editor and a map view (Figure~\ref{fig:interface}).
The JSON editor allows for the editing, saving, and loading of visualization specifications through an embedded JSON editor.
We use JSON for its conciseness and portability (\textbf{D2.2}).
%
%
As the user adds new views, layers, knots, plots, or changes parameters in the JSON file, the map view is updated.
If the user navigates the 2D or 3D environment and changes camera parameters, the new information will be saved in the JSON file.
The map view will also display the linked and embedded plots specified using Vega-Lite's grammar.
%

\highlight{When specifying \grammar{layer} elements, the user defines a path to the .utk file as the name of the layer. 
These references can only be made inside the \grammar{\text{layer}\textsubscript{in}} element of the \grammar{integration\_scheme}.
This is the only place in the specification where the user can make direct references to locally stored data. Otherwise, the data is only manipulated as \grammar{knots}.}
This allows for isolation between data and interface so that the interpreter can guarantee consistency of operations.


\noindent \textbf{Data interaction.}
%
%
\highlight{Interactions are supported on the map and on the 2D plots by specifying an \grammar{interaction} for \grammar{map} and \grammar{plot} elements, respectively.}
%
%
Selected entities in the 2D plots are highlighted on the map and vice versa.
This link between plots and physical elements is facilitated by the knots.
A reference to a knot is required when creating a new plot -- hence the data ingested to Vega-Lite is going to be dictated by how the knots are created.
For example, if a knot is aggregating data by ZIP area, the data passed to the plot will be at the same resolution. And if a bar chart is displayed, each bar element will have a $1:1$ mapping to a particular ZIP code. A selection in the chart will then highlight the appropriate area.
\subsection{Implementation \& analysis workflow}

\noindent \textbf{Implementation.}
%
%
The UTK framework has been developed using a client-server architecture.
\highlight{UTK's frontend uses React for user interfaces and WebGL for the 3D map component.
UTK's backend was implemented using Python. For reverse geocoding and to access OSM data, we use GeoPy and Overpass, respectively. Spatial joins and aggregations are performed using Geopandas' functionalities.}

%
%
%
%
%
%
%
%
%

\begin{figure}[t!]
    \centering
    \includegraphics[width=1\linewidth]{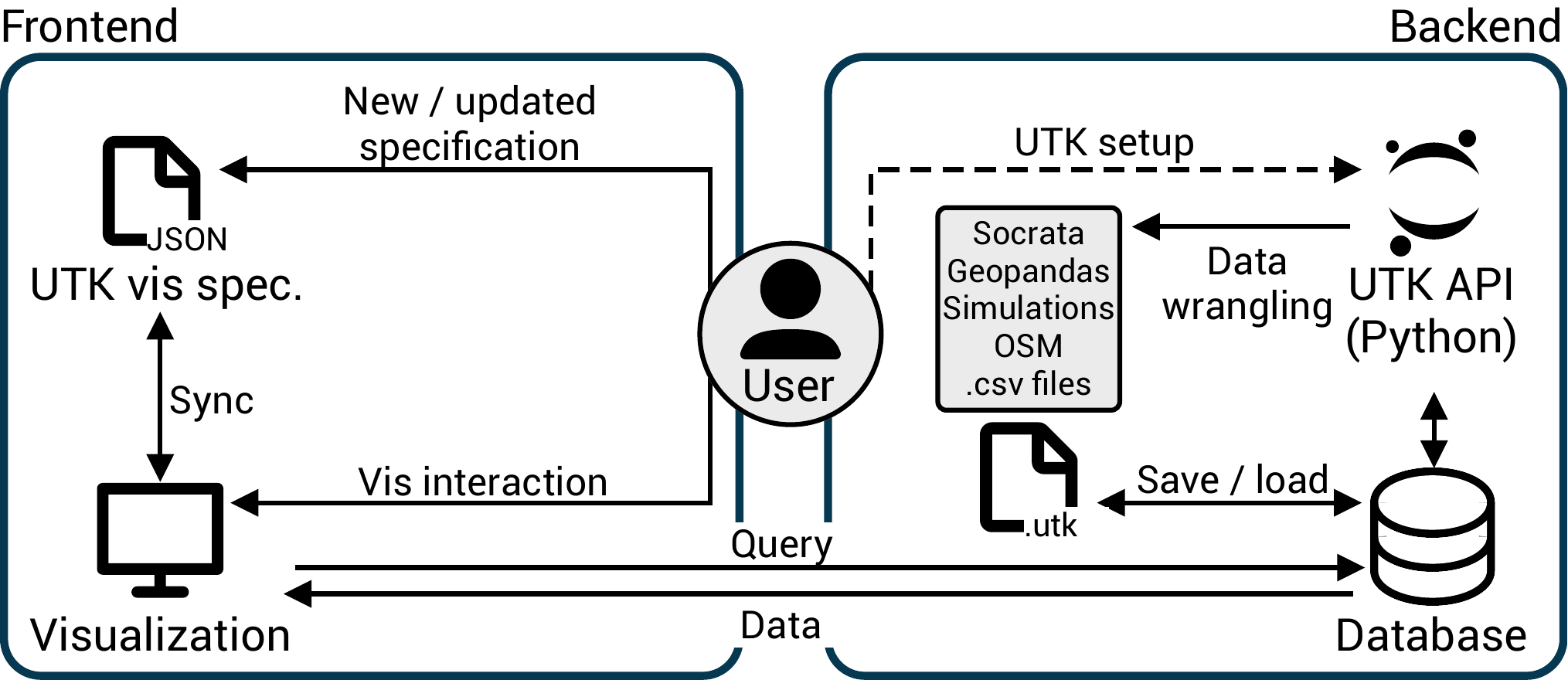}
    \caption{Overview of UTK's architecture and how the user interacts with the frontend and backend components of the framework.}
    \vspace{-0.6cm}
    \label{fig:system}
\end{figure}

\noindent \textbf{Analysis workflow.}
Figure~\ref{fig:system} presents a high-level overview of UTK's architecture and the expected ways that the user can interact with both UTK's frontend and backend.
%
%
If no data was previously created, in the first step, the user should set up the UTK environment (Figure~\ref{fig:system}~(right)).
This involves creating a new Jupyter Notebook and \highlight{specifying an area of analysis using UTK's Python API.}
After that, the physical data layers will be created and persisted as .utk files.
Even though the framework already offers a number of data functionalities, users can also make use of other libraries and load their output into the framework as long as the data adheres to UTK's formats.
Having a separation between the frontend and the backend enables users to go back and forth between UTK's Python API and the grammar visualization functionalities.
After the data is created, the user can visualize it using the frontend and use UTK's grammar to create the visualizations.
\highlight{When deploying to the web, users can optionally hide the JSON editor and only expose the map view.}
A tight connection between visualization and data through knots ensures that the user has enough flexibility to iteratively modify the data in Jupyter Notebooks and easily visualize the new data without having to worry about setting up the environment from scratch (\textbf{D3.2}).

\section{Evaluation}
\label{sec:evaluation}

To demonstrate UTK's capabilities, flexibility, and ease of use, we present examples motivated by real-world problems and inspired by previous collaborations with architects and urban planners.
%
%
%
Then, we report on a series of one-hour semi-structured interviews with five different urban experts, where we asked them their perspectives on UTK's usability, limitations, and potential extra features to be added.

\subsection{Example gallery}
\label{sec:useCases}

Next, we highlight examples showcasing how UTK can be used in different analysis workflows, motivated by the needs of urban domains.
%
%
%
%
%
%
%


\subsubsection{Example 1: Building energy efficiency}

In this use case, we are exploring buildings in Chicago's Near North Side area to identify locations in each building that are the most energy efficient in terms of receiving more shadows during summer than winter.
\highlight{As noted in previous studies~\cite{ichinose_impacts_2017, wang_sustainability_2021}, more shadows during summer can lead to cooler internal temperatures, and more sunlight during winter has the opposite effect. In both cases, shadows can contribute to a lower dependency on AC.}
%
%
%
\highlight{We start this analysis by using UTK's backend to create a physical layer with Chicago's buildings and twelve thematic layers with the accumulated shadows for a day in each month of the year. Creating these layers is simple, requiring less than 10 lines of Python code.}
\highlight{We then move to UTK's frontend and, using the JSON editor, we create twelve knots, each linking the physical layer with a thematic layer.}
\highlight{To model the trade-off between summer and winter shadows, we create a new knot that averages the accumulated shadows, with weights depending on the month (see Miranda et al.~\cite{miranda_shadow_2019} for the weights).}
Again, the concept of knots provides an easy abstraction for averaging the accumulated shadows.
%
%
Note that UTK's expressivity allows us to chain aggregations. We define a separate knot for each month's shadow and perform an operation over knots to get the final result, as shown in Figure~\ref{fig:teaser}~(d, top).

UTK also allows us to easily integrate 2D and 3D visualization to minimize occlusion problems.
\highlight{Figure~\ref{fig:teaser}~(d, bottom) highlights the specification of an \grammar{embedded} radial plot that displays the data distribution along the horizontal cross-section (\grammar{footprint}), with shades of blue denoting facades that have more positive than a negative shadow, according to our weighted average.}
\highlight{One could also use iterate over different types of plots, as shown in Figure~\ref{fig:plots}.}
With minimal effort in terms of navigation, it is possible to assess that the selected buildings have very different cross-section profiles.
By analyzing data at the building level, experts can gain insights into the unique characteristics of individual buildings and their surrounding environments, allowing them to develop targeted energy consumption strategies. Moreover, the analysis can reveal which floors and sides of each building have been exposed to more summer shadows on average.
Through the data analysis of building and land use, experts can also identify candidate areas for urban farming and green infrastructure~\cite{palliwal_3d_2021}, promoting more sustainable and resilient urban development.
%

\subsubsection{Example 2: Historic preservation}


Preserving historic landmarks is key to the preservation of cultural heritage. One of the important but often overlooked causes of building deterioration is the shadow cast by neighboring buildings~\cite{lee_building_2017}. 
%
In Boston, the shadow cast by new developments has long been the subject of public debate. 
%
In this case, we will use UTK for \emph{what-if} or alternate scenario testing. We have chosen the John Hancock Tower in Back Bay, Boston -- a 60-story skyscraper. We would like to measure the accumulated shadows cast by Hancock Tower on the 146-year-old Trinity Church and the Public Garden.
\highlight{We start by using UTK's backend to download OSM data for Boston and perform the shadow simulation for winter and summer. Next, using UTK's frontend, we create four knots, integrating physical and thematic layers for summer (with and without the tower) and winter (with and without the tower).}
%
%
Here, UTK's capabilities in computing differences between knots can help compute the amount of accumulated shadows cast by the tower.
Considering two knots with the computed shadow for the current and alternate states, we can simply create a new one and compute their difference:
\begin{grammarenv}
\noindent knot\textsubscript{diff}=(diff,(knot\textsubscript{s0},knot\textsubscript{s1}, nearest, knot\textsubscript{s0}-knot\textsubscript{s1}))
\end{grammarenv}

\begin{figure}[t]
    \centering
    \includegraphics[width=0.95\linewidth]{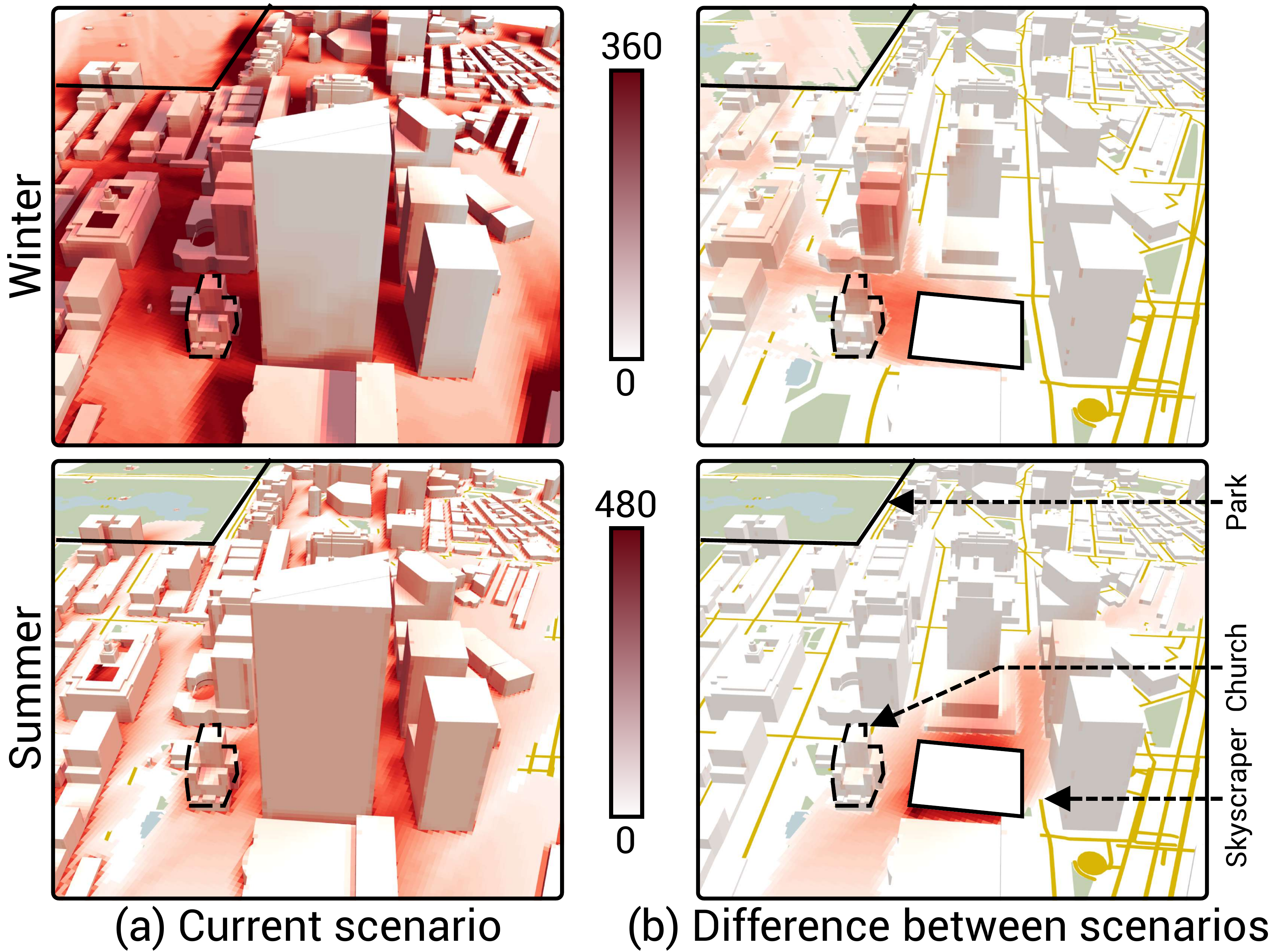}
    \caption{Evaluating shadows cast by a skyscraper in Boston. (a) The current shadow situation during winter and summer. (b) Amount of shadow cast solely by the skyscraper -- i.e., shadow that would disappear with the removal of the building. The color scales show the accumulation period (360 minutes during winter and 480 minutes during summer).
    }
    \vspace{-0.5cm}
    \label{fig:boston}
\end{figure}

Figure~\ref{fig:boston}~(a) shows the current state of accumulated shadows during summer and winter (the number of accumulated minutes is shown near the color scale). The right panel depicts the amount of shadow that would disappear with the removal of the tower; in other words, UTK enables us to easily extract the amount of accumulated shadow cast by this specific building. 
We can see from the figure that during summer, the orientation of the building resulted in a low amount of accumulated shadow cast over the church compared to the total accumulated shadow cast by other buildings in its vicinity. During winter, the eastern side of Trinity is more severely shadowed by the tower. 
As illustrated in Figure~\ref{fig:boston}~(b, top), during winter, the shadow cast by the Hancock Tower also spreads over the Public Garden, located miles away.

\subsubsection{Example 3: Neighborhood signatures}

Cities are often described by their most prominent features, such as cultural diversity and economic activities. 
%
Recognizing the unique characteristics of each neighborhood is crucial in planning and resource allocation. 
%
%
Given that, identifying patterns and understanding the dynamics between livability measures require analyzing datasets across scales~\cite{ferreira_urbane_2015}. 

%
%
In this example, we will use eight datasets from the NYC Open Data portal~\cite{noauthor_nyc_nodate} for neighborhood characterization: crime and noise reports; restaurant, parks, and subway locations; sky exposure; school quality reports and taxi pickups. 
\highlight{We start by using the backend to parse the CSV files from the identified datasets and create the thematic layers. In this step, we also use the backend to download OSM data for NYC and parse shapefiles with the neighborhoods.}
\highlight{Using the frontend, we then specify knots that aggregate thematic layers over physical ones.}
For example, to create thematic and physical layers:

\begin{grammarenv}
\noindent layer\textsubscript{noise}=(noise,coords),
layer\textsubscript{zip}=(zip,objs,coords)
\end{grammarenv}

Aggregations and joins can be performed through knots that link physical and thematic layers. We then create a knot with a single integration scheme between \grammar{layer\textsubscript{noise}} and \grammar{layer\textsubscript{zip}}, specifying that noise complaints within ZIP areas will be added together:

\begin{grammarenv}
\noindent knot\textsubscript{noise}=(noise2zip,(layer\textsubscript{noise},layer\textsubscript{zip}, contains, sum))
\end{grammarenv}

We can then specify a Vega-Lite parallel coordinate plot that visualizes the data from knots created for each of the previously mentioned dataset.
This creates the interface shown in Figure~\ref{fig:interface} (from a UTK specification with fewer than 200 lines).
The map shows a neighborhood in Manhattan with the most noise complaints. Looking at other indicators, we can see that it has a low sky obstruction rate, signaling a lower density neighborhood with relatively low access to public transit. 
Considering that it does not have a large number of taxi pickups and it is not a destination for food services, this elevated number of noise complaints seems alarming. 
Further investigation into the types of complaints is required to address the concerns of its residents. This insight was easily extracted from simple interactions with UTK, enabling users to spend more time on analysis and exploration. 

%

\subsubsection{Example 4: Tripping risk in Downtown Boston}

Sidewalk surface is among the most important factors in determining the risk of outdoor falls~\cite{twardzik_what_2019}. 
%
Slippery surfaces pose a major challenge to pedestrians navigating the outdoor environment in cold and snowy weather~\cite {chippendale_neighborhood_2015}.
In the absence of sunlight, and when the temperature suddenly drops, a transparent and slippery form of thick coating of ice, known as \emph{black ice}, can form on top of pavements. In general, the temperature of the surface and the type of surface material are key contributing factors to black ice formation~\cite{houle_winter_2008}. 

In this example, we use surface material data~\cite{hosseini_citysurfaces_2022} together with shadow accumulation data aggregated at the sidewalk level to calculate a risk measure for identifying places that can pose a higher risk of falls in cold seasons~\cite{hosseini_towards_2022}. 
Our risk measure is calculated using three factors of interest to urban accessibility experts: the percentage of bricks, the percentage of granite, and the accumulated shadow data for a day in December (winter solstice).
\highlight{UTK's backend was used to create a physical layer from a shapefile with sidewalk geometry and thematic layers from our shadow simulation and from a CSV file with the sidewalk material data.} 
%
We classified dangerous sidewalks as the combination of two knots, one with accumulated shadow and another one with the sidewalk material. We considered dangerous sidewalks as the ones with bricks and concrete, and more than half of the time under shadow during the accumulation period.
%
%
This can be easily represented through the operation:

\begin{grammarenv}
\noindent knot\textsubscript{danger} = (danger, (knot\textsubscript{shadow}, knot\textsubscript{mat},nearest,op))\\
op=((mat==`brick'||mat==`conc') \&\& shadow>0.5?0:1)
\end{grammarenv}

\begin{figure}[t!]
    \centering
    \includegraphics[width=0.9\linewidth]{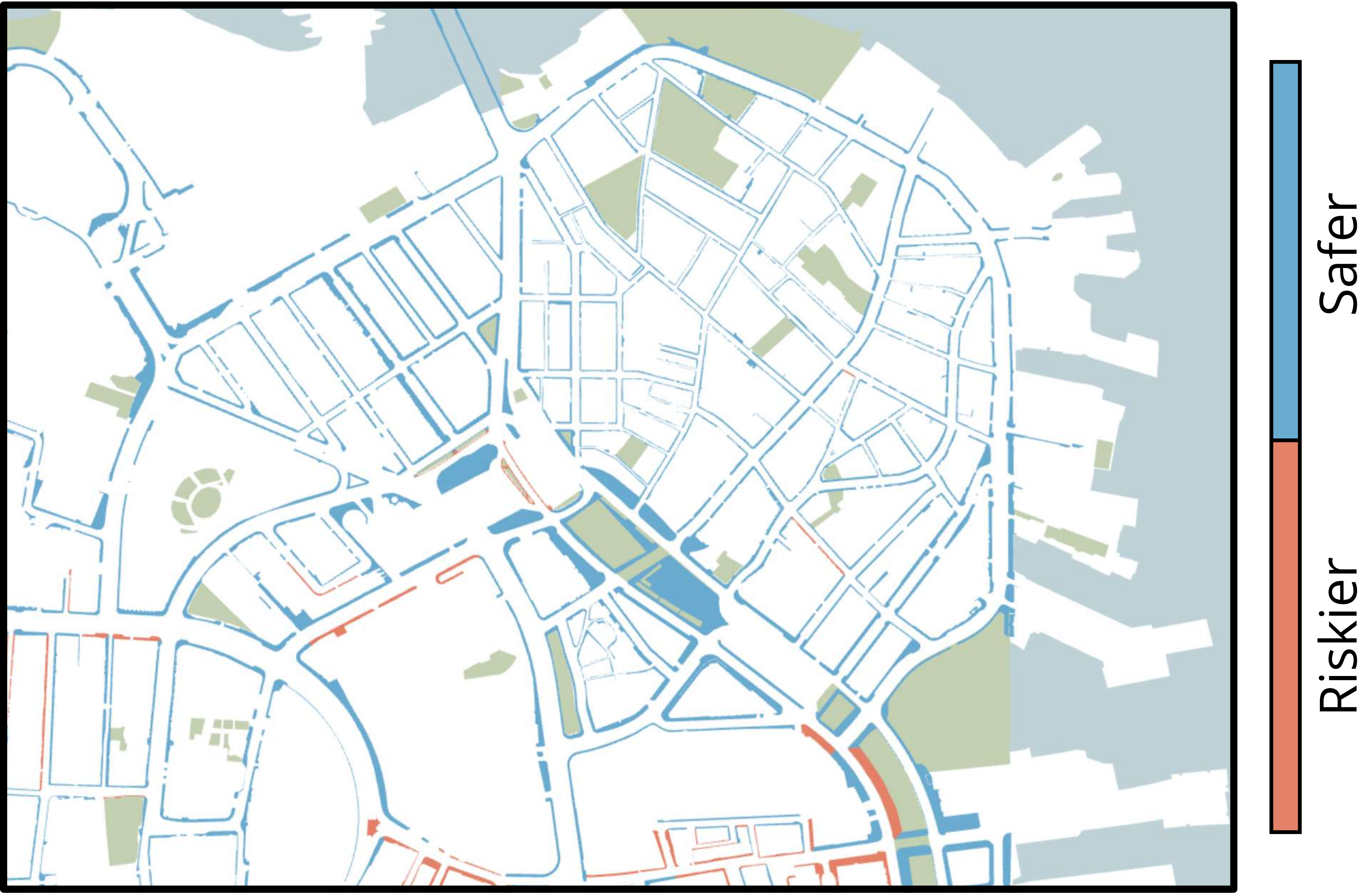}
    \caption{Knot's custom operator specifies whether a sidewalk is dangerous or not based on its surface material and shadow.}
    \vspace{-0.5cm}
    \label{fig:fallrisk}
\end{figure}

We chose Boston's downtown and North End neighborhoods for their historic fabric and because they are considered popular destinations. Figure~\ref{fig:fallrisk} shows our tripping risk map. 
%
UTK allows for easy iteration over multiple parameters through a simple combination of knots. The operation can be adjusted based on user's preferences and city-scale fall risk models \highlight{by simply editing the JSON specification}.

\subsection{Reflection on design goals}
\label{sec:reflections}

\highlight{In terms of extensibility, UTK provides the ability to create custom plots and integrate them into map-based visualizations. Our examples show two different possibilities (radial plot, parallel coordinates) of juxtaposed and embedded plots, satisfying \textbf{D1.1}.
Examples 1 and 4 highlight the use of knots to perform operations across layers (\textbf{D1.2}).
Example 4, specifically, goes toward tackling a design consideration recently highlighted by Saha et al.~\cite{saha_visualizing_2022}, i.e., facilitating the blending of datasets for multivariate analysis.
Throughout the examples, UTK abstracted many low-level steps involved in urban data analytics, namely downloading and parsing OSM data, joining data for visualization purposes, rendering, and navigation, satisfying \textbf{D1.3}.
From an implementation perspective, UTK's backend and frontend follow a modular approach, ensuring that new functionalities can be added without the need to modify the core code.}

\highlight{For reproducibility, the examples highlight UTK's support for standard formats, namely CSV files in example 3 and shapefiles in example 4, supporting \textbf{D2.1}. Also, rather than relying on cumbersome deployments, visualizations can be shared through small and concise JSON files (\textbf{D2.2}).}
\highlight{Considering its flexibility, UTK also provides a suite of data functionalities that facilitate porting use cases for other regions (\textbf{D3.1}). Examples 1 and 2 could be easily replicated in other cities just by changing the layer creation step. Similarly, new datasets could be added to example 3 or other assessment factors could be considered in example 4.
The examples also highlight analyses at different scales (sidewalk, building, and neighborhood).}
With UTK, visualizing the basic layers of a city requires only a few lines of Python code to import the data, along with a JSON file specifying the desired visualization. UTK abstracts all necessary data transformations, enabling users to focus on the analytical aspects of their work without worrying about the technical details (\textbf{D3.2}).

\subsection{Experts' feedback}
\label{sec:expertFeedback}

To gain domain perspectives into several aspects of UTK, we interviewed five urban experts from different specialties, roles, and levels of seniority.
None of the experts are co-authors in this paper.
\highlight{
We asked the experts to asynchronously interact with UTK by reading a quick start document and three tutorials highlighting different use cases.
Then, we proceeded to a semi-structured interview where the expert was asked questions regarding UTK's features, limitations, and usability.
%
}
%
%
%
We interviewed: two climate scientists with experience in climate and urban morphology based in North America (P1 and P2), an urban planner with experience in urban micro-climate based in Asia (P3), an architect and urban planner interested in the analysis of urban morphology (P4), a geographer and GIS expert that worked as the deputy secretary of urban planning for a large city in South America (P5).
%
%
P5 has an MSc degree, and the others have PhDs.
%
%

\noindent \textbf{Use cases.}
P1 and P2 were primarily interested in using UTK for heat vulnerability analysis. They mentioned it would be interesting to \emph{``integrate temperature, humidity, and shading for city-scale analysis and compare seasons.''}
They were also receptive to how UTK moves \emph{``away from static maps and enables alternate scenario planning, ideal for climate change resilience.''}
Finally, P1 mentioned how UTK \emph{``facilitates engagement not only across disciplines, but also across urban communities''}, as certain outcomes can \emph{``target users with different levels of expertise.''} 
P1 mentioned that the visual analysis could be \emph{``authored by an expert, but delivered to a community through a web portal, hiding the grammar from them.''}
One concrete example was the possibility to create \emph{``hourly temperature shadow maps for communities to minimize heat exposure.''}

P3 was responsive regarding the multi-scale capabilities offered by UTK, enabling \emph{``switching contexts and scales in a fast way''} and the \emph{``exploration of how different sides of buildings are exposed to sunlight.''}
In their practice, P3 uses temperature and sky view factor simulations to compute the pedestrian level of comfort. They mentioned that UTK's \emph{``capacity to aggregate data and perform a difference of the results is especially useful to assess the impact of different parameters and conditions of the simulations''}. 
%
%
P4 pointed out the potential for collaborative analysis using the framework: \emph{``researchers could easily share their visualizations instead of cumbersome GIS files.''} They mentioned that making the fast shadow accumulation computation available could lead to major use cases given that it's \emph{``easy to use and integrate into existing workflows.''}
Moreover, noise and property management were brought up as potential use cases.
P5 mentioned that the tool is ideal for analyzing urban morphology and that \emph{``3D makes it more attractive to users and a great tool for communication.''} They mentioned how, for urban land use decisions, \emph{``3D is important to assess possible new developments and communicate that to stakeholders.''}

\noindent \textbf{Usability \& adoption.}
We asked the urban experts how easy it would be to integrate UTK into their current practices.
They all mentioned that they see value in the tool.
P3 mentioned that, by adopting 2D and 3D metaphors, the framework can \emph{``cater to a broader audience, as opposed to just supporting one or the other.''}
\highlight{Regarding UTK's usability, P4 mentioned the need for more intuitive ways to interact with the physical layer: \emph{``it would be helpful to click on a building and interactively change its height.''}
P4 was also ambivalent regarding having access to the grammar, pointing out that \emph{``learning what JSON I can write takes time''}, but appreciated that they \emph{``don't have to work with too many tools and menus.''}}
P5 was also more cautious, saying that \emph{``it would require training to educate people on the grammar, with examples to showcase its use.''} But on the other hand, \emph{``GIS tools are complex and cumbersome, lightweight solutions would be very welcomed.''}
They mentioned that the \emph{``usability of current GIS tools is a problem and experts sometimes resort to limited toolboxes such as Google Maps / Earth for their vis.''} 
And that \emph{``even though data is being made available, visualizations are usually very simple, whatever is supported by Google Maps.''} 
P5 highlighted that \emph{``UTK could easily plug into a data lake to visualize bus data, with operational indicators for the city.''}
%
%
Moreover, it provides a \emph{``lightweight approach to visualize cities in 3D, good for both city management and communication.''}

%


\subsection{Comparison with existing tools}

\highlight{
In this section, we provide a comparison between UTK and popular GIS tools, namely ArcGIS and QGIS.
One of the drawbacks of these tools is the overwhelming number of operations and toolboxes.
%
For instance, ArcGIS Pro has more than 200 different operations in only one of its 41 toolboxes, each having its specific requirements for data and processing, and according to previous work, users often struggle to find the right one for their task~\cite{ziegler_need_2023}.
UTK, on the other hand, offers a concise specification that is easy to learn and modify. The grammar-based framework makes it possible to perform a wide range of visual analytic tasks with our grammar elements, offering a compelling alternative to traditional GIS tools. The ability to define custom links between layers summarizes a range of geospatial operators into a single functionality.

Another limitation of existing tools is the lack of support for reproducible and shareable formats in many of their toolboxes.
To reproduce an analysis done with those operations, a user needs access to the tool, which might require costly licensing and a cumbersome installation process.
%
%
In contrast, UTK saves the visualization and processing specification in human-readable JSON files. By building their own chain of operations, users can easily iterate designs, ensuring more transparency and reproducibility of their work. The self-contained nature of grammar-based visualizations allows for easier collaboration, sharing, and replication of visualizations.
ArcGIS and QGIS also have limited capabilities in creating plots and embedding them into physical layers~\cite{hu_usability_2017}.
UTK provides customizable plotting functionalities (through Vega-Lite), as well as the ability to manipulate plots with flexible integration options (e.g., embedded on building surfaces~\cite{mota_comparison_2022}). Beyond embedding plots, UTK directly links them to the underlying data source, enhancing interactivity and contextualization.
}

\section{Conclusion and future work}
\label{sec:conclusion}

This paper presents the Urban Toolkit, a flexible grammar-based declarative framework for urban visual analytics.
UTK streamlines many of the usual data tasks common in urban analyses and, as shown in our use cases, makes it possible to prototype complex analysis scenarios relatively easily\highlight{, in line with what has been recently highlighted by Saha et al.~\cite{saha_visualizing_2022} and Ziegler and Chasins~\cite{ziegler_need_2023}.}
We envision that our toolkit can popularize the process of authoring urban visual analytics systems, as well as facilitate user studies for a better understanding of the benefits of 2D and 3D visualizations in urban environments~\cite{mota_comparison_2022}.
%

While promising, our current implementation presents some limitations.
As shown in the examples throughout the paper and the accompanying video, UTK can leverage data concerning different layers from reasonably large portions of a high-density urban area \highlight{(e.g., Manhattan with approximately 3 million triangles)}.
%
%
However, high-resolution geometric data can also make matching operations between physical and thematic layers computationally expensive.
\highlight{As of now, our implementation uses Geopandas to perform spatial joins, a library that was not optimized for interactivity.}
We intend to investigate how to integrate more efficient algorithms and data structures to enable higher resolution physical models and thematic data\highlight{, as well as faster joins}~\cite{doraiswamy_interactive_2018}.
\highlight{Furthermore, UTK's support for temporal data is limited. In our current implementation, different knots can be used to aggregate data referring to different time ranges, but that is not scalable. We intend to investigate extensions to the grammar and the incorporation of timeseries data structures to better handle temporal data.
}
%
%

Since urban planning is tightly linked to simulations for scenario planning, we plan to incorporate other simulation capabilities, such as wind and noise, and make them available for researchers and practitioners through our API.
%
%
%
%
In addition, \highlight{we plan to conduct a controlled user study to evaluate ease-of-use}, as well as investigate the deployment of applications in different platforms, such as immersive environments~\cite{sicat_dxr_2019,butcher_vria_2021}.
\highlight{By making UTK public, we hope to pave the way for a more robust ecosystem of open tools for urban visual analytics.}

\section*{Acknowledgments}
We would like to thank the reviewers for their constructive comments and feedback. This study was partly funded by the University of Illinois' Discovery Partners Institute (DPI), CNPq (316963/2021-6), and FAPERJ (E-26/202.915/2019, E-26/211.134/2019).

\bibliographystyle{abbrv-doi-hyperref}

\bibliography{references}

\end{document}